
\input harvmac
\rightline{\vbox{\baselineskip12pt\hbox{CALT-68-1819}\hbox{hep-th/9209058}}}
\bigskip
\centerline{\bf Do Black Holes Destroy Information?\footnote{$^*$}{To appear
in the proceedings of the International Symposium on Black Holes,
Membranes, Wormholes, and Superstrings, The Woodlands, Texas, 16-18
January, 1992.}}\bigskip
\centerline{John Preskill}
\centerline{\it California Institute of Technology, Pasadena, CA 91125}
\bigskip
{\baselineskip=12pt{\narrower{I review the information loss paradox that was
first formulated by Hawking, and discuss possible ways of resolving it.  All
proposed solutions have serious drawbacks.  I conclude that the information
loss
paradox may well presage a
revolution in fundamental physics.}\smallskip}}
\bigskip
{\bf Introduction}.  Over 15 years ago, Stephen Hawking proposed that the usual
rules of quantum mechanics do not apply to a process in which a black hole
forms and then completely evaporates\ref\hawkA{S. W. Hawking, {\it Phys. Rev.}
{\bf D14} (1976) 2460.}.
If this proposal is correct, then we face the daunting task of
finding a new conceptual basis for all of physics.  Since Hawking's original
work, this issue has been much debated, but it has not been definitively
resolved.

When I began thinking seriously about black holes a few years ago, I was
inclined to dismiss Hawking's proposal as an unwarranted extrapolation from an
untrustworthy approximation.  It seemed to be based on the premise that no (or
hardly any) information about the body that collapsed to form the black hole
can be extracted from the thermal radiation that the black hole emits.  As best
I could tell, this premise was founded on the semiclassical calculation of
Hawking radiation\ref\hawkB{S. W. Hawking, {\it Comm. Math. Phys.} {\bf 43}
(1975) 199.}, in which all gravitational back--reaction effects are
neglected.  Rather than accept Hawking's remarkable (and radical) suggestion,
it seemed to me much more sensible (and conservative) to assume the validity of
quantum mechanics, and to try to understand the mechanism by which information
about the collapsing body gets encoded in the outgoing radiation.\foot{The
greatest champions of this viewpoint have been Page\ref\page{D. N. Page, {\it
Phys. Rev. Lett.} {\bf 44} (1980) 301.} and 't Hooft\lref\hooftA{G. 't Hooft,
{\it Nucl. Phys.} {\bf B256} (1985) 727.}\lref\hooftB{G. 't Hooft, {\it Nucl.
Phys.} {\bf B335} (1990) 138, and references therein.}\refs{\hooftA,\hooftB}.}
I was
hopeful that a detailed analysis of back reaction would reveal this mechanism.
But I also expected that the mechanism would be sufficiently  subtle and
enlightening as to notably deepen our understanding of fundamental physics.  I
was especially hopeful that light would be shed on the meaning of the
intrinsic black hole entropy.

As I have pondered this puzzle, it has come to seem less and less likely to me
that the accepted principles of quantum mechanics and relativity can be
reconciled with the phenomenon of black hole evaporation.  In other words, I
have come to believe more and more (only 15 years behind Hawking) that the
accepted principles lead to a truly
paradoxical conclusion, which means that these principles cannot provide a
correct description of nature.  This may be analogous to the conclusion,
derivable from classical physics, that the energy spectrum of black body
radiation diverges at short wavelengths.  Conceivably, the puzzle of black hole
evaporation
portends a scientific revolution as sweeping as that that led to the
formulation of quantum theory in the early 20th century.  Surely, this would be
the most exciting possible outcome of the puzzle.

If the currently accepted version of quantum theory really is fatally flawed,
then we must face the challenge of finding a new, self--consistent,
formulation of the fundamental laws that agrees with experiment.  So far, this
task has attracted surprisingly
little attention.  Perhaps the time has come to intensify the search.

\bigskip
{\bf The Paradox}.  Hawking's discovery of black hole radiance established a
deep and satisfying connection between gravitation, quantum theory, and
thermodynamics.  Particularly beautiful is the formula that
Hawking\ref\hawkC{S. W. Hawking, {\it Phys. Rev.} {\bf D13} (1976) 191.}
derived
(and Bekenstein\ref\bek{J. D. Bekenstein, {\it Phys. Rev.} {\bf D7} (1973)
2333; {\bf D9} (1974) 3292.} anticipated) for the intrinsic black hole entropy
\eqn\i{
S={1\over 4}~A~,}
where $A$ is the area of the event horizon in Planck units.  This elegant
result relegates the area theorem\ref\hawkD{S. W. Hawking, {\it Comm. Math.
Phys.} {\bf 25} (1972) 152.} of classical general relativity to a special
case of the second law of thermodynamics.

At the same time, black hole radiance raises some serious puzzles.  One puzzle
concerns the interpretation of black hole entropy.  In other contexts,
statistical--mechanical entropy counts the number of {\it accessible}
microstates that a system can occupy, where all states are presumed to occur
with equal probability.  If a black hole has no (or very little) hair, the
nature of these microstates is obscure.  Eq.~\i\ invites us to construe the
horizon as a quantum membrane with about one degree of freedom per Planck unit
of area\hooftB, but a more concrete conception of these degrees of freedom
remains
elusive.

More distressing is a serious paradox raised by Hawking\hawkA.  In his
semiclassical
calculation of black hole radiance, Hawking had found that the emitted
radiation is {\it exactly} thermal.  In particular, the detailed form of the
radiation does not depend on the detailed structure of the body that collapsed
to form the black hole.  This is because the state of the radiation is
determined only by the geometry of the black hole {\it outside} the horizon,
and the black hole has no hair that records any detailed information about the
collapsing body.  (While the semiclassical approximation used by Hawking is not
exact, it is quite plausible that the emitted radiation really is only weakly
correlated with the state of the collapsing body.  The key constraint comes
from causality---once the collapsing body is behind the horizon, it is
incapable of influencing the radiation.\lref\waldA{R. Wald, ``Black Holes,
Thermodynamics, and Time Reversibility,'' in {\it Quantum Gravity 2}, ed. C. J.
Isham, R. Penrose, and D. W. Sciama, Oxford (1981).}\foot{See, for example,
\waldA.})

That the radiation outside the black hole is in a mixed (thermal) state is in
itself neither surprising nor disturbing.  After all, the region outside the
horizon is only part of a quantum system, and there are correlations between
degrees of freedom (quantum fields) that are accessible outside the horizon,
and
the inaccessible degrees of freedom that are behind the horizon.  It is because
of these correlations that the radiation detected by observers outside the
horizon is in a mixed state.

But suppose that the black hole continues to evaporate until it disappears
completely.  Now the radiation {\it is} the whole system.  And so it seems that
 an initially pure quantum state, by collapsing to a black hole and then
evaporating completely, has evolved to a mixed state.  In other words, even if
the initial quantum state were precisely known, we cannot predict with
certainty what the final quantum state will be; we can only assign
probabilities to various alternatives.  This is the information loss paradox.
The paradox is that if we try to analyze the evolution of a black hole using
the usual principles of relativity and quantum theory, we are led to a
contradiction, for these principles forbid the evolution of a pure state to a
mixed state.
It is a familiar fact of life that information is often lost {\it in
practice}.\lref\zurek{W. Zurek, {\it Phys. Today} {\bf 44} (1991) 36.}\foot{For
a recent review, see \zurek.}
Here something essentially different is being claimed, that information is
actually destroyed {\it in principle}.

\bigskip
{\bf Information regained?}
Hawking concludes that the usual rules of quantum mechanics cannot apply in all
situations, which means that the fundamental laws of physics must be
reformulated.  Is there really no way to avoid this extraordinary conclusion?
Let us examine some of the alternatives.\lref\giddings{S. B. Giddings, {\it
Phys. Rev.} {\bf D46} (1992) 1347.}\lref\harvstrom{J. A. Harvey and A.
Strominger, ``Quantum Aspects of Black Holes,'' Enrico Fermi Institute Preprint
(1992).}\foot{For other recent discussions, see \giddings\ and \harvstrom.}
\bigskip
\leftline{\it 1. Can the information come out with the Hawking radiation?}
To a pragmatic physicist, the most likely place for the information to be
hiding is in the Hawking radiation emitted by the black hole.  After all, if we
throw a volume of the encyclopedia into the sun, then for all practical
purposes,
information is destroyed.  But we don't really believe that the information
about the initial quantum state has been lost in principle.  Even as the
encyclopedia burns beyond recognition, all of the information that it carried
presumably becomes stored in subtle and intricate correlations among the
radiation quanta emitted by the sun, or correlations of the emitted quanta with
the internal state of the sun.  Information is lost in practice because we are
unable to keep track of all these correlations.

Now suppose that we try to read the encyclopedia by measuring the sunlight.
Even if we measure the properties of the emitted radiation to arbitrary
accuracy, not much information comes out at first.  The radiation is in a
nearly thermal mixed state because it has complicated correlations with the
internal state of the sun (which we are not measuring).
But if we wait long enough for the ``sun'' to settle down to its unique quantum
ground state, and stop radiating, then only correlations among the emitted
quanta can carry information.  There are certainly plenty of ways to encode
information in correlations among quanta emitted by the system at different
times.  If we measure those correlations to sufficient accuracy (and we know
the precise initial quantum state of the sun, before we threw in the
encyclopedia), we can recover
the encyclopedia.

So why should a black hole be fundamentally different?  The pragmatic viewpoint
holds that, in a similar way, the Hawking
radiation emitted by a black hole seems at first to be in a mixed state\page.
But
by the time the black hole has radiated away most of its mass, there are
detailed and subtle correlations between the quanta emitted early and the
quanta emitted later on.  These correlations, in principle, carry all of the
information about the quantum state of the initial collapsing body.

Since a black hole has no (or little) hair, the pragmatic viewpoint challenges
us to
explain how the black hole manages to record the information about the quanta
that it has already emitted, so that it is able to induce these correlations.
But there is a sharper way (impressed on me by Lenny Susskind) of expressing
why the pragmatic view is implausible, and difficult to reconcile with
causality\lref\susskind{L. Susskind, private communication.}\lref\sussthor{L.
Susskind and L. Thorlacius, ``Hawking Radiation and Back Reaction,'' Stanford
Preprint SU-ITP-992-12 (1992).}\refs{\susskind,\sussthor}.  On the spacetime of
an evaporating black hole, it is possible to
draw a single spacelike slice that crosses most of the outgoing Hawking
radiation, and {\it also} crosses the collapsing body, well inside the
(apparent) horizon.  (We can also choose this slice to stay far from the
singularity, in regions of low curvature, so we are confident that we know the
causal structure reliably.)

\def\ket#1{|#1\rangle}
Now, we know that if the outgoing radiation by itself is in a nearly pure
state, it must {\it not} be strongly correlated with the state of the body
inside the horizon.  The trouble arises because, if the quantum state outside
the horizon is really uncorrelated with the state inside the horizon, then it
follows from the principle of superposition that the state inside the horizon
must be a unique state that carries no information at all.  The argument goes
like this:  Let $\{\ket{i}\}$ denote a basis for the initial quantum state of
the collapsing body, and take the extreme view that each of these states
evolves to a state on the  spacelike slice constructed above, such that the
radiation and the collapsing body are {\it completely} uncorrelated; we have
\eqn\ia{\ket{i}\longrightarrow\ket{i}_{\rm inside}\otimes\ket{i}_{\rm outside}}
---the final state is the tensor product of a pure state inside the horizon and
a pure state outside.  But we may also consider a superposition of these basis
states, which evolves as
\eqn\ib{\sum_i c_i\ket{i}\longrightarrow\sum_i c_i\left(\ket{i}_{\rm
inside}\otimes\ket{i}_{\rm outside}\right).}
In general, the state inside and outside will be correlated, {\it unless} all
of the states $\ket{i}_{\rm inside}$ are actually the {\it same} state.  So the
radiation will always be in a pure state only if the body is in a {\it unique}
state.
(This is reminiscent of how the ``sun'' settled down to its unique ground state
in the example cited above.)
More generally, if the radiation state is nearly pure, then the body's state
must be nearly unique.
We conclude that, if the information really
propagates out encoded in the Hawking radiation, then there must be a mechanism
that strips away (nearly) all information about the collapsing body as the body
falls through the apparent horizon (and long before the body reaches the
singularity).  In the lively imagery of Susskind\susskind, a mysterious force
must bleach the encyclopedia as it tumbles into the black hole, removing the
message that it contains.  It is hard to imagine any reasonable way to achieve
this,
because to a freely falling observer the apparent horizon is not a very special
place.

If bleaching of the information at the horizon does not occur, then macroscopic
violation of causality seems to be required to transport the information from
the collapsing body to the outgoing radiation.  At the very least, the
semiclassical picture of the causal structure must be highly misleading.

\bigskip
\leftline{\it 2. Can the information be retained by a stable black hole
remnant?}
The pragmatic view is that small corrections to the leading semiclassical
theory build up over time, so that by the time the black hole has radiated away
most of its mass, most of the information has been recovered.  An advantage of
this scenario is that we can hope to understand and analyze it without invoking
Planck--scale physics.  Most other proposed ways of escaping information loss
are based on speculations about how a Planck--mass black hole behaves.

If semiclassical theory is not misleading, then the Hawking radiation emitted
by a large black hole reveals little information about the collapsing body.  If
information is not lost, this must mean that the information is retained inside
the black hole.  When the black hole has evaporated down to the Planck size,
the standard semiclassical theory of black hole evaporation is surely no longer
applicable, as spacetime is subject to violent quantum fluctuations on this
scale.  We can not be sure what happens next without a deeper understanding of
quantum gravity.

Perhaps quantum gravity effects halt the evaporation process, so that a stable
black hole remnant is left behind.  At first sight, this seems to resolve the
information loss paradox, because all of the information about the initial
collapsing object can in principle reside in the remnant.  But upon further
reflection, the cure may be worse than the disease.  Since the initial black
hole could have been arbitrarily massive, the remnant must be capable of
carrying an arbitrarily large amount of information (about $
M^2/M_{\rm Planck}^2$ bits, if the initial mass was $M$).  This means that
there must be an infinite number of species of stable remnant, all with mass
comparable to $M_{\rm Planck}$.

It seems hard to reconcile this sort of infinite degeneracy with the
fundamentals of quantum field theory, that is, with analyticity (causality) and
unitarity\hooftA.  The coupling of the remnants to hard quanta might be
suppressed by
form factors, but the coupling to soft quanta (wavelength $>> ~L_{\rm Planck}$)
should be well-described by an effective field theory in which the remnant is
regarded as a pointlike object.  Then the coupling to soft gravitons, say,
should be determined only by the mass of the remnant, and should be independent
of its internal structure, including its information content.  We should be
able to use this effective field theory to analyze, for example, the emission
of Planck--size remnants in the evaporation of a large black hole.  For each
species, the emission is suppressed by a tiny Boltzman factor
$\exp({-\beta_{\rm Hawking}M_{\rm remnant}})$.  But if there are an infinite
number of species, the luminosity is nonetheless infinite.

The emission of Planck--size remnants in the evaporation of a large black hole
is merely an example of a soft process in which heavy particles can be
produced, a process that is expected to admit an effective field theory
description.  If such processes really have infinite rates (as would be
expected if the are an infinite number of Planck--mass species), then these
infinities will inevitably infect other calculated processes, as a consequence
of unitarity.  These infinities would be quite malevolent---they would destroy
the consistency of the theory.

So if stable remnants are really the answer, an effective field theory
description of the coupling of the remnants to soft quanta cannot be
valid---the coupling must depend on the hidden information content of the
remnant.  Banks {\it et al.}\ref\banks{T. Banks, A. Dabholkar, M. R. Douglas,
and M. O'Loughlin, {\it Phys. Rev.} {\bf D45} (1992) 3607; T. Banks and M.
O'Loughlin, ``Classical and Quantum Production of Cornucopions at Energies
Below $10^{18}$ GeV,'' Rutgers Preprint RU-92-14 (1992).} have recently offered
a particularly vivid explanation of how this might be possible.  In their
picture, the information that resides in a black hole remnant is contained in a
long, narrow throat that is attached onto spacetime.  Production of remnants is
heavily suppressed because it is necessary to add a large volume (the volume of
the throat) to the background spacetime, and this process requires a large
Euclidean action.

For a number of reasons, the arguments in \refs{\banks} are not very
convincing.
The primary motivation underlying the suggestion that a black hole remnant has
a long throat comes from studies of dilaton gravity, which arises as a
low--energy limit of string theory.  The extreme magnetically charged black
holes in this theory\lref\gibbons{G. Gibbons and K. Maeda, {\it Nucl. Phys.}
{\bf B298} (1988) 741.}\lref\garfinkle{D. Garfinkle, G. Horowitz, and A.
Strominger, {\it Phys. Rev.} {\bf D43} (1991) 3140.}\refs{\gibbons,\garfinkle}
really do have infinitely long throats (if the length is measured by the
``string metric'' that determines how strings propagate on the background).
But while the throat of the magnetically charged black hole is threaded with
magnetic flux that prevents the throat from pinching off, it is unclear what
would prevent the throat of an uncharged black hole from pinching off.  If the
pinch--off occurs,  the information stored in the throat would be lost to a
``baby universe'' disconnected from our own universe (see {\it (5.)} below).
Furthermore, the infinite volume that potentially allows the extreme hole to
store vast quantities of information may be illusory.  Dilaton gravity becomes
strongly coupled far down the throat of the extreme black hole.  The total
throat volume in the weakly--coupled region is actually quite modest.  Thus, to
argue persuasively that a dilatonic black hole harbors a large amount of
information, one must perform a nonperturbative analysis that currently seems
intractable.  Finally, especially since semiclassical methods are not really
applicable, it is quite difficult to calculate production rates for the extreme
holes, or to otherwise support the contention that the production rate is at
odds with the effective field theory viewpoint.

Giddings\giddings\ recently suggested a variation on the stable remnant idea,
that a
black hole that harbors a lot of information actually stops evaporating when it
is still large compared to $L_{\rm Planck}$.  The more information, the larger
the remnant.  So the number of species less than a specified mass $M$ is always
finite, and the contributions of remnants to soft processes can be heavily
suppressed.  It strikes me that this suggestion is at least as peculiar as the
idea that effective field theory cannot be applied to the tiny remnants.  The
odd thing is that there must be arbitrarily large black holes that emit no
Hawking radiation, contrary to semiclassical theory.  This failure of
semiclassical theory must occur even though the curvature at the horizon is
arbitrarily small.

Another displeasing feature of the remnant idea (in either of the two forms
above) is that it leaves us without a
reasonable interpretation for the Hawking--Bekenstein entropy.
If information is really encoded in the Hawking radiation, then it seems to
make sense to say that $e^{S(M)}$ counts the number of accessible black hole
internal states for a black hole of mass $M$.  But if the information stays
inside the black hole, then the number of internal states has nothing to do
with the mass of the black hole.
Indeed (if the remnants are Planck size), we can prepare a black hole of mass
$M$ that holds an arbitrarily large amount of information, by initially making
a much larger hole, and then letting it evaporate for a long time.  Thus, the
number of possible internal states for a black hole of mass $M$ must really be
infinite.
The beautiful edifice of black hole thermodynamics then seems like
an inexplicable accident.

(If a black hole really destroys information, then the interpretation of the
intrinsic entropy must be somewhat different, but perhaps still
sensible.  The black hole entropy measures the amount of {\it inaccessible}
information.  As the black hole evaporates, the entropy is transferred to the
outgoing radiation.  The entropy of the radiation does not result from coarse
graining---the mixed density matrix characterizing the radiation is really an
{\it exact} description of its state.)

Note that if we reject the idea of stable black hole remnants, there is a very
important consequence---there can be no exact continuous global symmetries in
nature.\lref\wheeler{J. A. Wheeler, unpublished.}\lref\bekB{J. D. Bekenstein,
{\it Phys. Rev.} {\bf D5} (1972) 1239.}\lref\zeldo{Ya. B. Zeldovich, {\it Sov.
Phys. JETP} {\bf 45} (1977) 9 ({\it Zh. Eksp. Teor. Fiz.} {\bf 79} (1977)
18).}\foot{That the ``baryon number'' of a black hole is ill--defined was first
emphasized by Wheeler\wheeler\ and Bekenstein\bekB. That the complete
evaporation of a black hole would transcend global conservation laws may have
been first stressed by Zeldovich\zeldo.} Suppose that $Q$ is a putative
conserved charge, and that $m>0$ is the
mass of the particle with the smallest mass-to-charge ratio.  (We'll
take its charge to be one.)  By assembling $N$ particles, we can create
a black hole with charge $Q=N$ and mass $M$ of order $Nm$; if $N$ is
large enough, we have $M>>M_{\rm Planck}$, so that semiclassical theory
can be safely applied to this black hole.  In fact, we can make $M$ so
large that the Hawking temperature is small compared to the masses of
all charged particles.  Then the black hole will radiate away most of
its mass in the form of light uncharged particles, without radiating
away much of its charge.  At this point, there is no way for the
evaporation of the black hole to proceed to completion without violating
conservation of $Q$; there is no available decay channel with charge
$Q=N$ and a sufficiently small mass.  The only way to rescue the
conservation law is for the black hole to stop evaporating, and settle
down to a stable remnant that carries the conserved charge.  Since this
doesn't happen in semiclassical theory, it seems that we are forced
either to Planck--size remnants, or the surprising breakdown of
semiclassical theory for large black holes envisioned by Giddings.  And
there would be an infinite number of species, because $N$ could take any
value.  If we accept the objections to the existence of an infinite
number of remnant species, then, we must accept the consequence that the
conservation law is violated.

This is an unusual kind of anomaly.  There is a conservation law that is exact
at the
classical level, but is spoiled by quantum effects.  Since the black hole
``forgets'' the value of the charge that it consumes, one may wonder whether
loss of information is unavoidable in theories that suffer from this anomaly,
theories in which the conservation law is violated ``only'' by processes
involving black holes.  I don't think that we are forced to this conclusion.
It is at least a logical possibility that
all of the information about the initial collapsing body is actually preserved
as in {\it (1.)} above.  That charge is not conserved (contrary to our initial
expectation) need not imply that information is destroyed.

Note that this argument for nonconservation breaks down if there are massless
particles that
carry the conserved charge. It also
does not apply (or at least, is not totally convincing) for
{\it discrete} global symmetries---for example, a $Z_n$
symmetry, where $n$ is not too large.

\bigskip
\leftline{\it 3. Can all of the information come out ``at the end?''}
When I said that ``the information comes out with the Hawking radiation'' under
{\it (1.)} above, I meant that, after most of the the mass of the black hole is
radiated away, the state of the radiation that has been emitted is not really
thermal, but is instead nearly pure.  Another logical possibility is that the
radiation remains truly thermal until much later (as the semiclassical theory
indicates).  Finally, when the black hole  evaporates down to the Planck size,
and semiclassical theory breaks down, information starts to leak out;
it is encoded in correlations between the thermal  quanta emitted earlier, and
the quanta emitted ``at the end.''

But if the black hole was initially very big, so that the amount of information
is very large, then the information can not come out suddenly.  The final stage
of the
evaporation process must take a very long time\lref\aharonov{Y. Aharonov, A.
Casher, and S. Nussinov, {\it Phys. Lett.} {\bf B191} (1987)
51.}\lref\carlitz{R. D. Carlitz and R. S. Willey, {\it Phys. Rev.} {\bf D36}
(1987)  2336.}\refs{\aharonov,\carlitz}.
To get an idea how long it must take, we should count the number of
quantum states that are available to the Planck-energy's worth of radiation
that is emitted in the last stage.  These quanta all have wavelengths
that are much larger than the size of the evaporating object, so it is an
excellent approximation to suppose that they
all occupy the lowest partial wave.  Thus, for the purpose of counting
states, the problem reduces to a one--dimensional (radial) ideal gas.

Actually, the same is true to a
reasonable approximation for a big black hole, since the emitted quanta
have wavelength comparable to the size of the hole.  As a warm up, let's
consider the case of a big black hole first, and check that the
Hawking--Bekenstein entropy counts the number of radiation states from
which the black hole can be assembled.  If the mass of the black hole is
$M$, then the radiation state from which it formed must contain energy
$M$ inside a sphere with radius
comparable to the Hawking evaporation time
$t_{\rm Hawking}\sim M^3$.  (I am now using units with $M_{\rm Planck}=1$).
The
entropy $S$ of a one-dimensional ideal gas  with with energy $E$ and
``volume'' $L$ is, in order of magnitude,
\eqn\ii{S^2\sim EL~.}
So for $E\sim M$ and $L\sim M^3$, we find $S\sim M^2$, the Hawking--Bekenstein
entropy.

(By the way, it is interesting to ask how the above analysis is modified if
there are $\nu$ different species of massless radiation, with $\nu>>1$.  Then
the entropy scales like $S^2\sim \nu EL$, but the Hawking time decreases like
$L\sim M^3/\nu$.  So  we see that $\nu$ drops out of the entropy\lref\thorne{W.
H. Zurek and K. S. Thorne, {\it Phys. Rev. Lett.} {\bf 54} (1985)
2171.}\refs{\hawkC,\thorne}, and we can
begin to understand how the black hole entropy can be a universal quantity,
independent of the details of the matter Lagrangian.)

Now let's ask what the volume of a one--dimensional ideal gas would have to be,
if the gas has the same entropy as above, but energy $E\sim 1$.  Or in other
words, how much would the gas have to expand adiabatically to cool down to
$E\sim 1$.  Evidently, it would need to expand by the factor $M$, so that
$L\sim M^4$.  If it takes a time $t_{\rm remnant}$ before the long--lived
remnant finally disappears, then the radiation emitted during this time
occupies a sphere of radius $L\sim t_{\rm remnant}$; we thus
obtain an upper bound\carlitz\
\eqn\iii{t_{\rm remnant}{\
\lower-1.2pt\vbox{\hbox{\rlap{$>$}\lower5pt\vbox{\hbox{$\sim$}}}}\ }M^4~.}
This bound is saturated if the final radiation is equilibrated---that is, if it
is able to occupy nearly all of the states that are available in the allotted
time.  Of course, the decay of the remnant might actually take much longer,
but it has to take at least this long.

Another way to say what is going on is that the remnant must emit about $S\sim
M^2$ quanta to reinstate the information.  Since the total energy is of order
one, a typical quantum has energy $M^{-2}$ and wavelength $M^2$.  Further, to
carry the required information, these quanta must be only weakly correlated
with one another.  This means, roughly speaking, that they must come out one at
a time, as non-overlapping wave packets.  Since the time for the emission of
each quantum is $M^2$, and there are $M^2$ quanta, the total time is $M^4$.

If the information comes out at the end, then, the scenario is that a black
hole with initial mass $M$ evaporates down to Planck size in time $M^3$, but
the time for the Planck--size remnant to disappear is much longer (at least
$M^4$.)  The trouble is that, since $M$ can be arbitrarily large, there must be
Planck--size black hole remnants that are arbitrarily long lived, even if no
species is absolutely stable.  If there are an infinite number of species with
mass of order the Planck mass, all with lifetime greater than googolplex, then
we have all the same problems as if the remnants were absolutely stable.

For the sake of logical completeness, I'll note a variation on the
Giddings\giddings\
idea about massive remnants---namely massive metastable remnants.  Perhaps the
evaporation of a black hole that harbors a great deal of information departs
significantly from semiclassical theory while the hole is still large compared
to Planck size, and it starts to emit quanta that have wavelength much longer
than the naive thermal wavelength.  The more information, the larger the black
hole when this starts to happen.  Then we would have an infinite number of
long-lived species, but only a finite number with mass below a given energy,
which might be acceptable.  But, we again would face the challenge of
understanding how semiclassical theory can fail so badly for very large black
holes.

\bigskip
\leftline{\it 4. Can the information be encoded in ``quantum hair?''}
An important ingredient in the information loss puzzle comes from the black
hole uniqueness theorems of classical general relativity.  It is because the
geometry outside the horizon is insensitive to the detailed properties of the
collapsing body that the Hawking radiation is uncorrelated with the state of
the collapsing body, in the leading semiclassical theory.  As we have noted,
one who holds the position that information is encoded in the Hawking radiation
is challenged to explain how corrections to semiclassical theory enable the
black hole to store an accurate record of how it was formed and what it has
already radiated.

The main conceptual point concerning ``quantum hair'' on black
holes\lref\bowick{M. J. Bowick, S. B. Giddings, J. A. Harvey, G. T. Horowitz,
and A. Strominger, {\it Phys. Rev. Lett.} {\bf 61} (1988) 2823.}\lref\krauss{L.
M. Krauss and F. Wilczek, {\it Phys. Rev. Lett.} {\bf 62} (1989)
1221.}\refs{\bowick,\krauss} is that
there are additional possibilities for hair that are missed in the  analysis of
black hole solutions of the classical field equations.  This enables the black
hole to record more information than we would naively expect, information that
influences the Hawking radiation in a calculable way\ref\CPW{S. Coleman, J.
Preskill, and F. Wilczek, {\it Mod. Phys. Lett.} {\bf A6} (1991) 1631; {\it
Phys. Rev. Lett.} {\bf 67} (1991) 1975; {\it Nucl. Phys.} {\bf B378} (1992)
175.}.  The moral is that the
``no-hair principle'' has limitations, and we should be cautious about drawing
sweeping conclusions from it.

On the other hand, the discovery of quantum hair seems at first sight to offer
us little guidance concerning the information loss problem.  The type of
quantum hair that has been analyzed in detail is associated with charges that
can be detected
by means of the Aharonov--Bohm effect.  Such quantum numbers arise only in
theories with special matter content.  Furthermore, even in theories that have
many varieties of quantum hair, it is possible to make a black hole that
doesn't carry any, and in that case
the Aharonov--Bohm effect doesn't enable us to find out much about the internal
state of the black hole.

The first objection above, that quantum hair arises only in theories with
special matter content, is not so compelling.  Indeed, a very exciting
possibility is that the effort to avoid information loss will lead us to a very
special class of theories, or even a unique one (perhaps superstring theory).
Still, it is hard to
imagine that quantum hair (of the Aharonov--Bohm type) can really resolve the
paradox.  It seems that the idea would have to be that there are an infinite
number of exactly conserved charges (associated with an infinite number of {\it
unbroken} gauge symmetries), so that measuring values of all the charges would
suffice to uniquely specify the internal state of an arbitrarily large black
hole.  These conservation laws would have to be quite different than the
conservation laws that we usually think about.  For example, suppose I make a
black hole by allowing $N$ hydrogen atoms to collapse.  If all the information
about the initial state is to be encoded in quantum hair, then it seems that
each possible state of the atoms must be in a distinct superselection sector!
The values of the conserved quantities are not at all what I would expect if I
tried adding together the charges of the individual atoms; they depend in a
highly nonlocal way on how the atoms are patched together (in flagrant
violation of cluster decomposition).  The conservation laws put exceedingly
powerful constraints on the evolution of the system, so powerful that it is
hard to understand how they have escaped notice in our low--energy experiments.
 If quantum hair really enables a black hole to retain vast amounts of
information, the challenge is to explain how local quantum field theory seems
to describe low--energy physics so well.

There is a claim\ref\nano{J. Ellis, N. E. Mavromatos, and D. V. Nanopoulos,
{\it Phys. Lett.} {\bf B267} (1991) 465; {\bf B272} (1991) 261; {\bf B284}
(1992) 27, 43.} that string theory provides just the sort of nonlocal
conservation laws that are required, in the form of ``$W$-hair,''
but I don't understand this idea well enough to give a useful appraisal.

\bigskip
\leftline{\it 5. Can the information escape to a ``baby universe?''}
Perhaps the most satisfying ``explanation'' for the loss of information in
black hole physics was offered by Dyson\ref\dyson{F. Dyson, Institute for
Advanced Study Preprint (1976), unpublished.}, Zeldovich\zeldo,  and
Hawking\lref\hawkE{S. W. Hawking, {\it Phys. Rev.} {\bf D37} (1988) 904; S. W.
Hawking and R. Laflamme, {\it Phys. Lett.} {\bf B209} (1988) 39.}\lref\hawkF{S.
W. Hawking, {\it Mod. Phys. Lett.} {\bf A5} (1990) 145,
453.}\refs{\hawkE,\hawkF}.  Their picture, described
in (rather misleading) classical language, is that quantum gravity effects
prevent the collapsing body from producing a true singularity inside the black
hole.  Instead the collapse induces the nucleation of a closed ``baby
universe.''  This new universe carries away the collapsing matter, and hence
all detailed information about its quantum state.  The baby universe is
causally disconnected from our own, and so completely inaccessible to us; we
have no hope of recovering the lost information.  Yet there is a larger sense
in which information is retained.  The proper setting for quantum theory, in
this picture, is a ``multiverse'' which encompasses the quantum--mechanical
interactions of all of the universes that are causally disconnected at the
classical level.  To the ``superobserver'' who (unlike us) is capable of
perceiving the state of the whole multiverse, no information is lost; it is
merely transferred from one universe to another.
In a more correct quantum--mechanical language, black holes produce
correlations between the state of the parent universe and the state of the baby
universe, and it is because of these correlations that both the parent and the
baby are described as mixed quantum states.

Still, it provides us little solace that the superobserver can understand what
is going on.  We want to know how to describe physics in the universe that we
have access to.  In this regard, it is quite important to observe that, since
the baby universe is closed, the energy that it carries away is precisely zero.
 Its energy (and momentum) being precisely known, its position in spacetime is
completely undetermined.  Thus, the baby universe wave function is really a
global quantity in our universe, with no spacetime dependence.  As
Coleman\ref\coleman{S. Coleman, {\it Nucl. Phys.} {\bf B307} (1988) 867.} and
Giddings and Strominger\ref\gidstrom{S. B. Giddings and A. Strominger, {\it
Nucl. Phys.} {\bf B307} (1988) 854.}
emphasized (in a somewhat different context than black hole physics), this
means the baby universe Hilbert space has a natural basis, such that different
elements of the basis correspond to different superselection sectors from the
perspective of our universe.  In each superselection sector, the baby universe
state is a unique pure quantum state, and it follows that our universe is also
described by a pure state.  Mixed states arise only if we commit the unphysical
act of superposing the different superselection sectors.

The baby universe idea, then, seems to lead us to the following picture\hawkF:
When
a pure state collapses to form a black hole, and then evaporates, it evolves to
a pure state.  This state is predictable in the sense that if we perform the
experiment many times with the same initial state, we always get the same final
state.  But the result of the experiment might not be predictable from the
fundamental laws of physics; it might depend on what superselection sector we
happen to reside in.  (The exception would be if there is a principle, a ``big
fix,'' that picks out a unique sector.)  There may be many, many
phenomenological parameters that we need to measure before we can predict
unambiguously how a black hole with initial mass $M$ will evaporate,
conceivably as many as $e^{S(M)}$.

Not only is this conclusion disheartening, but we are still left without a
satisfactory resolution of the information loss puzzle.  Once we have
measured all of the relevant parameters, and can make predictions, we still
long to
learn the {\it mechanism} by which the black hole remembers the initial state
so that it knows how to evaporate.  This leads us back to contemplate {\it
(1.)--(4.)} above.

\bigskip
{\bf Outlook}.  The claim that black holes destroy information seems like a
wild leap, until we examine the alternatives.  All of the possibilities listed
here seem to require rather drastic revision of cherished ideas about physics.
Perhaps we are just being really stupid, and a crucial insight that has eluded
us will allow everything to fall into place.  But it seems increasing likely to
me that it is as hopeless to reconcile relativistic quantum mechanics with
black hole
evaporation as it would have been to understand the spectrum of black body
radiation using classical physics.  The information loss paradox may be a
genuine failing of 20th century physics, and a signal that we must recast the
foundations of our discipline.

The case against the self--consistency of relativistic quantum theory is not
yet airtight.  We can hope to make the case stronger, even without achieving a
much deeper understanding of quantum gravity at short distances.  In some
respects, the hypothesis of an infinite variety of stable (or very long--lived)
black hole remnants may be the most conservative proposed way of avoiding
information loss.  The arguments against remnants can be sharpened and
generalized, or perhaps promising loopholes will be found.

The hypothesis that information is encoded in the outgoing Hawking radiation
can also be fruitfully investigated.  As noted above, it should be possible to
address the issue without considering the intricacies of quantum field theory
at large spacetime curvature.  Nor is it necessary to consider so complex a
process as the gravitational collapse of many quanta to form a macroscopic
black hole.  Instead, the question of information loss can be probed by
studying scattering of a single quantum off of an extreme black
hole\ref\pres{J. Preskill, P. Schwarz, A. Shapere, S. Trivedi, and F. Wilczek,
{\it Mod. Phys. Lett.} {\bf A6} (1991) 2353.}.  Extreme black holes emit no
Hawking radiation, and in suitable models are absolutely stable objects.  If
quantum coherence is maintained, then the scattering should be described by an
$S$-matrix.  But if black holes destroy information, then no $S$-matrix should
exist.  Furthermore, an extreme black hole with a large charge is a big object,
so that the scattering process does not seem to involve very--short--distance
physics.

In a seminal paper, Callan, Giddings, Harvey, and Strominger\ref\callan{C. G.
Callan, S. B. Giddings, J. A. Harvey, and A. Strominger, {\it Phys. Rev.} {\bf
D45} (1992) 1005.} pioneered the analysis of scattering off of an extreme black
hole in dilaton gravity, using a (1+1)--dimensional semiclassical approximation
that systematically includes gravitational back--reaction effects.  (The
dilaton is
invoked so that one can plausibly argue that the quanta absorbed and emitted by
the black hole are all in the $S$-wave; thus, a truncation to an effective
(1+1)--dimensional field theory is reasonable.)  Their work stimulated much
subsequent investigation over the past few months, which has been reviewed in
\harvstrom.  The analysis of this system turns out to be more involved than
initially envisioned, and cannot be completely carried out within the domain of
validity of semiclassical methods.  Still, further progress is likely to be
achieved, and to provide new insights.  A sufficiently thorough analysis might
convincingly demonstrate that information is really lost during the scattering
process.

If we conclude that quantum mechanics must be overturned, how are we to
proceed?  Hawking\lref\hawkG{S. W. Hawking, {\it Comm. Math. Phys.} {\bf 87}
(1983) 395.}\refs{\hawkA,\hawkG} suggested a new dynamics that specifies
the evolution of a density matrix, rather than a wave function.  His proposal
was sharply, and cogently, criticized by Banks, Peskin, and Susskind\ref\BPS{T.
Banks, M. E. Peskin, and L. Susskind, {\it Nucl. Phys.} {\bf B244} (1984)
125.}.  They
emphasized the difficulty of reconciling loss of quantum coherence with other
principles, such as locality and conservation of energy.  Very loosely
speaking, loss of coherence can be modeled by coupling a quantum system to a
source of random noise.  But noise tends to heat a system up.  If fluctuations
at the Planck scale destroy coherence very efficiently, then the Planck-scale
``noise'' ought to produce a lot of quanta at the Planck frequency. This
doesn't seem to happen, nor could such a failure of energy conservation be
easily accommodated in general relativity.

Part of the challenge before us is to find a ``phenomenological'' description
of information loss.  We need a generalization of quantum field theory, one
that preserves the successful low--energy predictions, yet can accommodate loss
of coherence (or acausal propagation) at some level.\lref\ellisA{J. Ellis, J.
S. Hagelin, D. V.
Nanopoulos, and M. Srednicki, {\it Nucl. Phys.} {\bf B241} (1984)
381.}\lref\ellisB{J. Ellis, N. E. Mavromatos, and D. V. Nanopoulos, ``Testing
Quantum Mechanics in the Neutral Kaon System,'' CERN Preprint CERN-TH.6596/92
(1992).}\lref\waldB{R. Wald, {\it Phys. Rev.} {\bf D21} (1980)
2742.}\foot{Phenomenological limits on loss of information are discussed in
\refs{\ellisA,\ellisB}.  Some issues of principle are addressed in
\refs{\waldA,\waldB}.}  Theories of this type might be very restricted in
form.  Banks {\it et al.}\BPS\ took the pioneering steps toward demonstrating
this.\lref\sred{M. Srednicki, ``Is Purity Eternal?'' Santa Barbara Preprint
UCSB-92-22 (1992).}\foot{For a more recent discussion, see \sred.}

One's devout wish is that experiment can guide us, as it guided Planck and his
followers.  Perhaps unexpected clues about the new physics will be uncovered,
or already have been without being recognized.  Meanwhile, it is not so
unrealistic to hope to make real progress via pure thought.  Anyway, we don't
have much choice.

\bigskip
{\bf Acknowledgments}.  This report was heavily influenced by
discussions with participants in the  workshop on Quantum
Aspects of Black Holes at the Aspen Center for Physics.  The evolution
of my thinking about black hole physics has been
facilitated by many colleagues, especially Tom
Banks, Sidney Coleman, Steve Giddings, Stephen Hawking, Alex Ridgway, Andy
Strominger, Lenny Susskind, Kip Thorne, Sandip Trivedi, and Frank Wilczek.
This work was supported in part by the US Department of Energy under
Contract No. DE-AC03-81-ER40050.
\listrefs
\bye